\documentclass[11pt,twoside]{article}
\usepackage{asp2010}
\usepackage{epsf}
\usepackage{psfig}
\usepackage{lscape}

\begin{document}
\title{E-accessible Astronomy Resources}
\author{Eva Isaksson}
\affil{Helsinki University Library, Kumpula Campus Library, 
PO Box 64, 00014 University of Helsinki, Finland}

\begin{abstract}
Making online resources more accessible to physically challenged
library users is a topic deserving informed attention from astronomy
librarians. Recommendations like WCAG 2.0 standards and
section 508, in the United States, have proven valuable, and some vendors are already making
their products compliant with them. But what about the wide variety
of databases and other resources produced by astronomy information
professionals themselves? Few, if any, of these are currently compliant
with accessibility standards. Here we discuss some solutions to
these accessibility challenges.

\end{abstract}

\section{Introduction}

A fair number of astronomers and astronomy students have a physical challenge. 
It is our responsibility to learn the basics of accessibility to be able
to help our library patrons to gain access to things that they need for their
studies and work.

Astronomy is often seen as a very visual science. After all, its origins lie in looking 
at the skies. Hence, it is a common belief that you need to use your sight to be able 
to study astronomy. This is strictly not true. In reality, we have been
using assistive technologies -- telescopes, sensors, computers -- for a long
time now to gain access to data that the human eye does not see unaided. 
Visual information is coming to us as large streams of bytes.

The modern astronomer is hardly bound by physical limitations.
One can produce solid research sitting  comfortably in front of one's personal computer.

There are many examples of physically challenged individuals who have
made successful careers in science. Those who have seen the movie 
\textit{Contact} based on Carl Sagan's novel are familiar with the blind
astronomer who is listening to radio signals instead of watching them
on the screen. His character is based on a real scientist, Dr. D. Kent Cullers.\footnote{http://www.seti.org/Page.aspx?pid=411} There are other
success stories -- in fact, too many to enumerate here.

\section{What Does Accessibility Look Like?}

But, you ask, isn't the sheer amount of information a major hindrance to 
those who cannot browse it easily? Yes, it is -- to some degree. 
Electronic textual materials provide both a possibility and a challenge 
for those with low vision.  In theory, it is possible for almost anyone to access 
online information, but in practice, this requires know-how and proper tools.

Plenty of assistive technologies exist to overcome hindrances. 
The Daisy standard\footnote{\texttt{http://www.daisy.org/}} for digital 
talking books has been an important tool for making electronic texts easy to 
browse. 

Not all hindrances are in the visual domain. Imagine an elderly astronomer
who has the full use of his or her intelligence, but whose hands are
shaking, and who might have some difficulty with pointing a mouse when
navigating a webpage and filling out search forms. 

It is a challenging task for librarians and information specialists to make 
our services and search forms accessible to people with a diversity of abilities 
so that they can do the research necessary for building careers as active contributors
in their chosen fields of research.

\begin{figure}[!ht]
\plotone{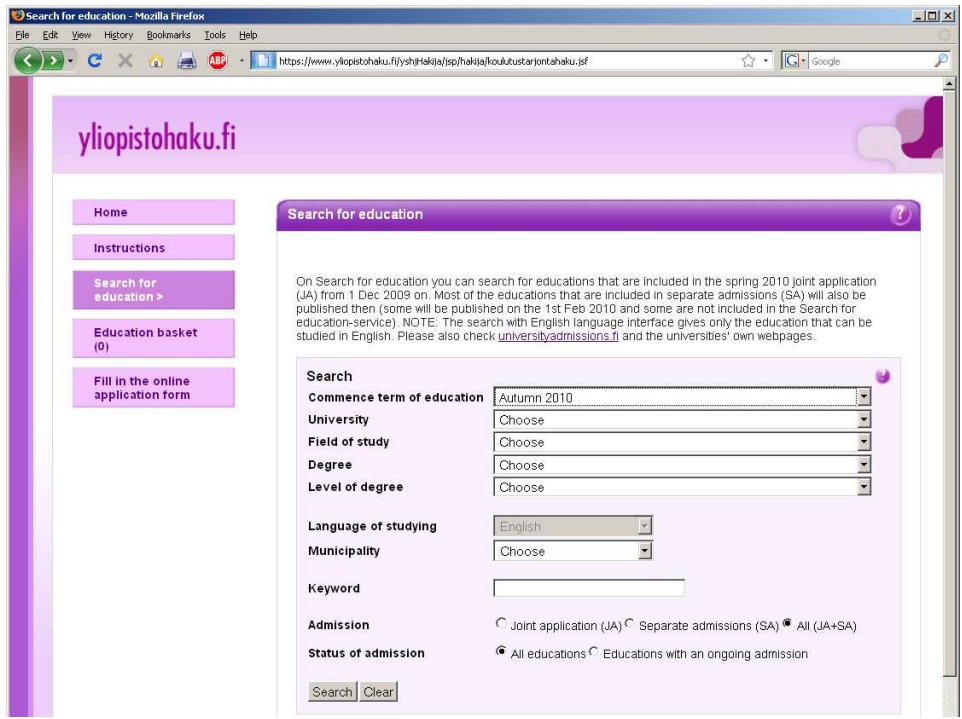}
\caption{This web page looks exactly the same before and after accessibility
changes. The only changes are on the HTML coding level, affecting the functionality 
but not the layout.}
\end{figure}

But what does accessibility look like? There is a pervasive myth that it looks 
boring. This is strictly not true. Accessible design should be functional enough, not just pretty. With proper HTML code and other techniques, we can make the text compliant with technological aids.
If the HTML coding is poor, a document may be impossible to open with such aids or it
could be impossible to navigate the text.

The author of this paper was involved with an university-wide accessibility 
project that was undertaken by  the University of Helsinki in 2005--2006, 
with a follow up in 2008--2009. It was recognized that accessibility  must 
cover not only our physical surroundings, but also the online environment as well.

In spring 2009, we noticed that the new national online system for applying 
for university education was not accessible to blind students. The system
was provided by the Finnish Ministry of Education, and we challenged them to 
fix it. To our big surprise, they did, working in collaboration with us 
and the Finnish Federation of the Visually Impaired.

Figure 1 shows a page from the application system. It looks exactly the same both 
before and after accessibility changes were made. Differences can be seen on 
the coding level, but otherwise one cannot tell the old version from the new one by visual inspection alone. The change has resulted in a major functional
improvement. The old version could not even be opened with assistive technology, 
and blind students could not use it. Now they can.

\section{Standards and How to Apply Them}

Accessibility needs some muscle to drive it. It is not just about good people doing 
good deeds -- it is also about ensuring that everyone has access to things that
matter to them. We need guidelines and standards, preferably with legislation to back 
them up.

In the United States, Section 508 of the Rehabilitation
Act regulates purchases made with federal funding. It is about ``access to and
use of information and data that is comparable to that provided to others.''
A market for accessible products helps big publishers to take accessibility 
into account. When a publisher has a large enough number of customers who need to 
buy  accessible products, they will be motivated to sell accessible products.

We also need strong standards. The World Wide Consortium has updated its Web
Content Accessibility Guidelines (WCAG) -- version 2 dates back to 2008.\footnote{W3 Web Content Accessibility Guidelines (WCAG ) version 2 (2008)\\ 
\texttt{http://www.w3.org/TR/UNDERSTANDING-WCAG20/}}
This new version of WCAG is meant to be a practical tool, evidenced by its three levels of accessibility: 

\begin{itemize}
\item A : minimum
\item AA : medium
\item AAA : as accessible as possible
\end{itemize}

You will find a good WCAG2 checklist online.\footnote{\texttt{http://webaim.org/standards/wcag/checklist}}
The ideal thing to do would be to make your website as accessible as
possible, but in practice you need to read the guidelines and 
identify the accessibility level best suited to serving your users.

Let's look at a concrete example by applying an A-level guideline to 
an existing search form. The guideline states: ``Form inputs have associated 
text labels or, if labels cannot be used, a descriptive title attribute.''

Let's look at a part of an ADS search 
form\footnote{\texttt{http://adsabs.harvard.edu/abstract\_service.html}} 
with its original coding. This piece of code is from the section which 
requires  an object for selection.

\vskip 0.2in

\begin{quote}
\noindent
\small\texttt{
<input name="obj\_req" value="YES" type="checkbox">\\
Require object for selection}
\normalsize
\end{quote}

\vskip 0.2in

Let's add some more coding (in boldface). Rather than just a checkbox, we now have a  \textit{text label}.\footnote{\texttt{http://www.webaim.org/techniques/forms/controls.php}} 

\vskip 0.2in

\begin{quote}
\noindent
\small\texttt{
<input id="obj\_req" \textbf{name="obj\_req"} value="YES" type="checkbox">\\
\textbf{<label for="obj\_req">}Require object for selection\textbf{</label>}}
\normalsize
\end{quote}

\vskip 0.2in

Figure 2 shows what has changed. The text label in question has been highlighted. 
It is no longer necessary to hit the small checkbox -- it is enough if you 
just click the associated text. This makes the box much easier to check.

\begin{figure}[!ht]
\fbox{
\plotone{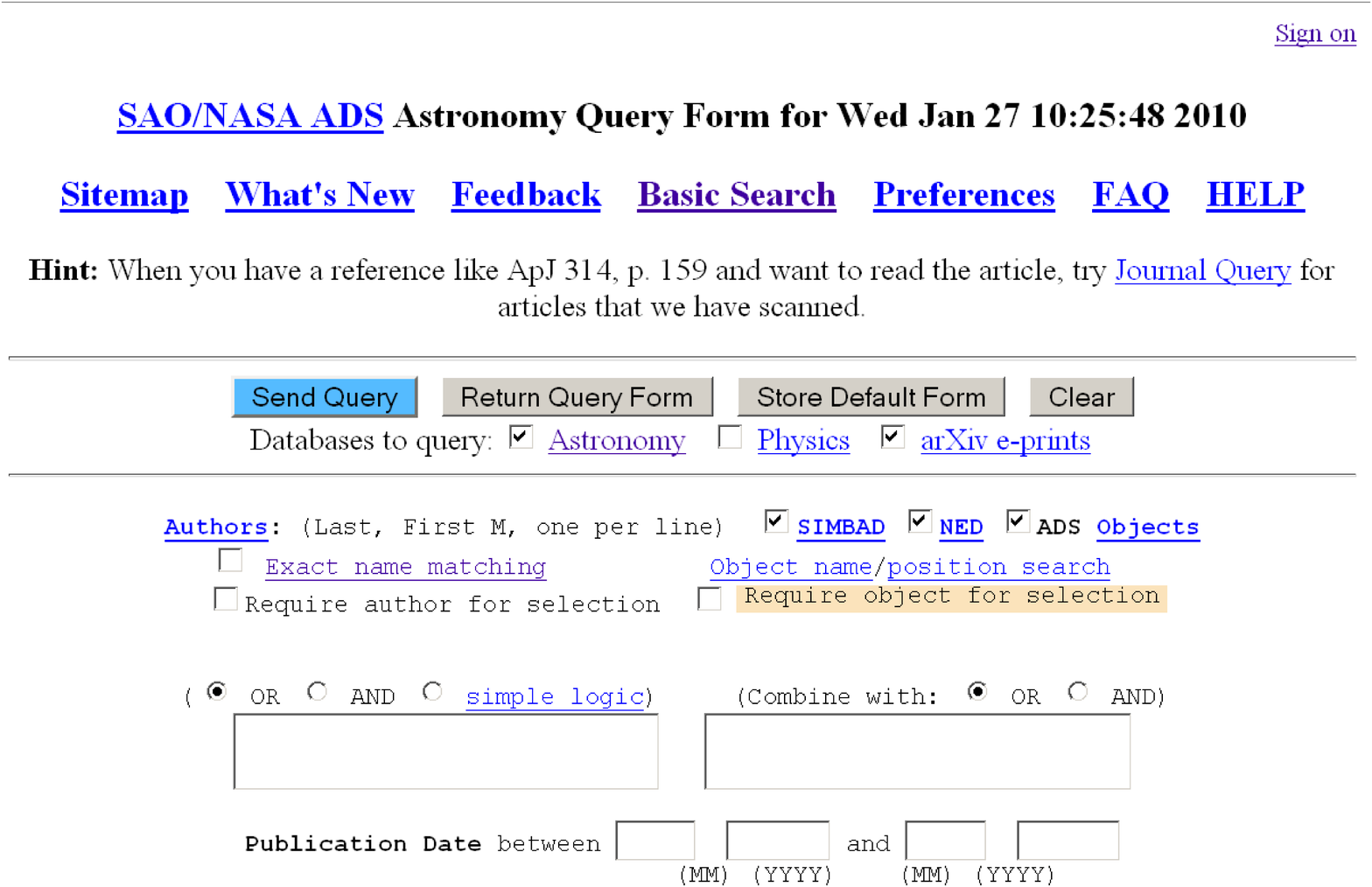}}
\caption{A slight change in the code could make it possible to click
the title of a checkbox (marked here with a different color) to check it}
\end{figure}

\section{Structure is the Key}

You can do clever things with HTML. There are however many other formats to 
consider: PDF, Flash, and office products, to name just a few.

No matter what the material at hand, it needs structure above all else. Otherwise, a blind person who tries to read a text has to read everything from beginning to end and is not able to navigate to a chapter or a footnote. Even PDF which used to be an accessibility nightmare can now 
boast of a structure to make it more accessible -- it's called tagged PDF.

\begin{figure}[h]
\begin{center}
\fbox{
\includegraphics[width=8cm]{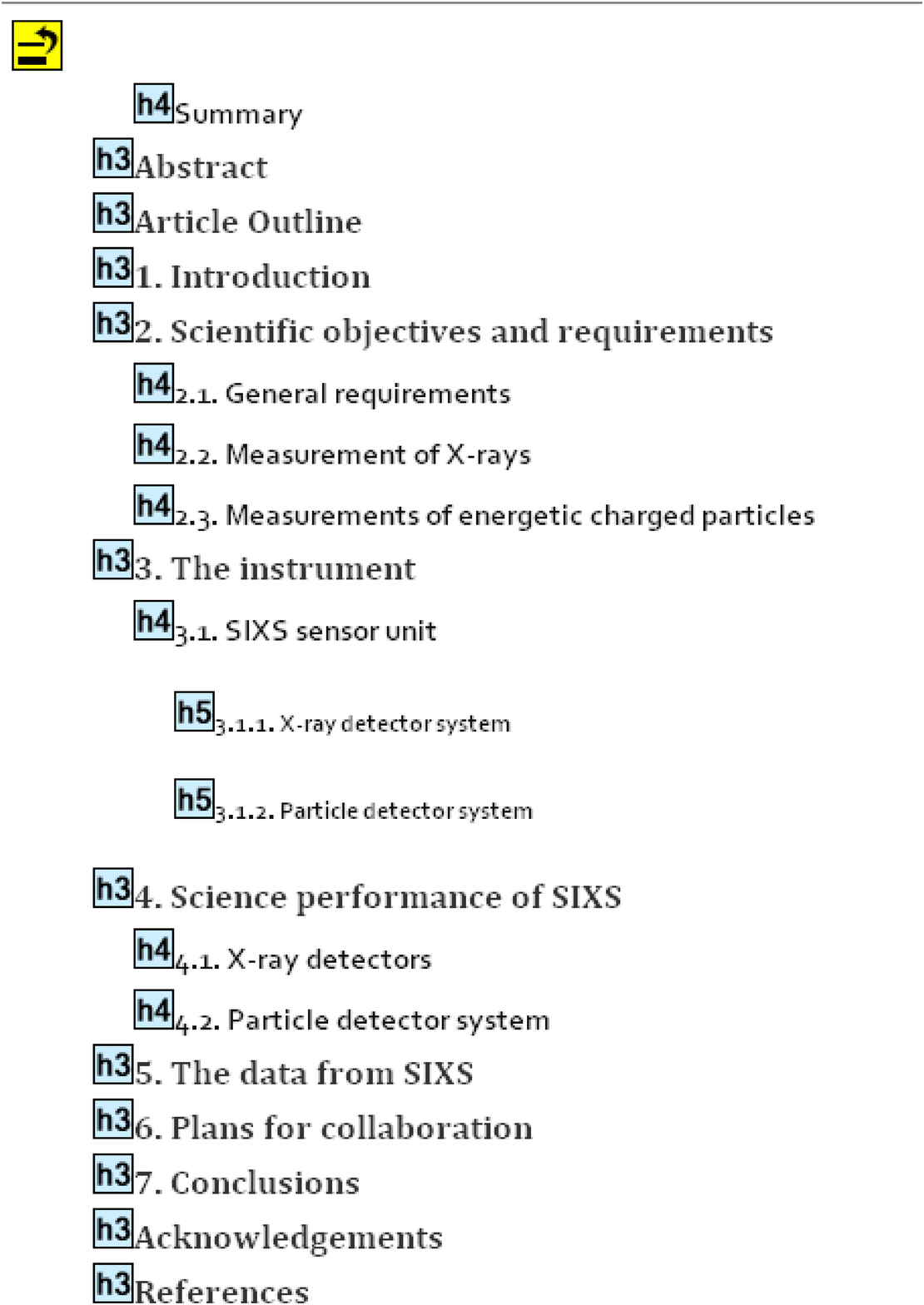}}
\end{center}
\caption{The structure of an astronomy paper revealed by the WebAIM tool.}
\end{figure}

As a general guideline, no matter what kind of document you are writing, 
you will need to stick to structure. Do you use subtitles that are bold
and in a different font? Please, use proper titles instead and use 
styles to control the fonts and such.

Let's take a peek at an HTML page that has structure. There are tools
to make the structure visible. The box in Figure 3 has been done with a 
WAVE toolbar\footnote{\texttt{http://wave.webaim.org/toolbar}}. 
This example is taken from \textit{Planetary and Space Science}. A good 
amount of structure has been revealed.  The HTML structure of  \textit{Earth, Moon and Planets}, shows next to nothing.  Its only structure is a references header, ``h2 References.'' There is no subtitle structure at all that you can jump to.

Most publishers make their electronic materials available in PDF format. Usually, 
those files are without any structure. Figure 4 shows the Acrobat Reader results
of an accessibility quick check -- there is no structure.

\begin{figure}[!ht]
\begin{center}
\includegraphics[width=9cm]{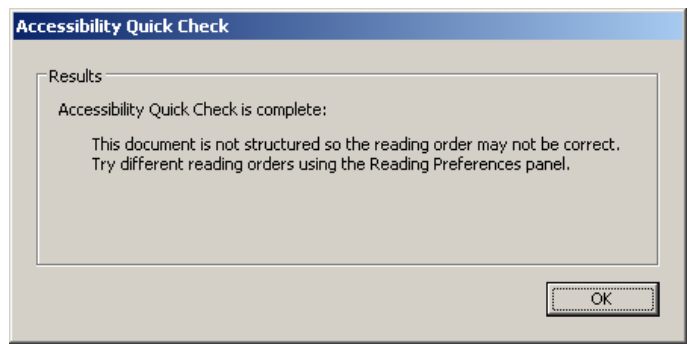}
\end{center}
\caption{Adobe Reader accessibility check tool warning about missing structure.}
\end{figure}

\section{Big Publishers and Small}

What is the current situation with different astronomy publishers and journals?
Table 1 shows accessibility elements for a selection of publishers based on inspection of a few papers published in 2009 by University of Helsinki astronomers. We asked some questions about
the basic properties of each paper. Is there HTML fulltext? Does it have structure? 
And does the PDF have structure? If not, are there at least PDF bookmarks?

You can see that these results leave a lot to hope for. The only
consistently good results are from \textit{Planetary and Space Science}, which is
published by Elsevier. Unfortunately, however, not all Elsevier 
products are equally accessible.

\begin{table}[!ht]
\caption{Quick review on the presence of accessibility elements in some astronomy journals}
\smallskip
\begin{center}
{\small
\begin{tabular}{llcccc}
\tableline
\noalign{\smallskip}
Title & Publisher &  HTML & HTML & PDF &  \\  
& &  fulltext & structure & structure & bookmarks \\  
\noalign{\smallskip}
\tableline
\noalign{\smallskip}
Astronomy \& Astrophysics & EDP Sciences & yes & ok & no & yes \\ 
Astrophysical Journal & IoP  & yes & none & no & yes \\ 
Monthly Notices R.A.S. & Wiley & yes & none & no & no  \\  
Astron. Nachrichten  & Wiley & no  & -- & no & no \\ 
Planetary Space Sci. & Elsevier & yes & ok & yes & yes \\ 
Earth, Moon \& Planets & Springer & yes & none & no  & yes \\ 
\noalign{\smallskip}
\tableline
\end{tabular}
}
\end{center}
\end{table}

Elsevier was the winner of this brief check. It has been making 
some efforts to increase accessibility of its products, which sets a good 
example for other big publishers.\footnote{See 
\texttt{http://www.info.sciencedirect.com/using/access\_article\_display/accessibility/} 
for Elsevier's ScienceDirect accessibility statement.} \citet{Isakssont2007} 
have inspected the overall accessibility compliance and practices of major database 
vendors, Elsevier included.

Even if major publishers are making some progress, it is not enough. 
There are also smaller publishers, and beyond that there are institutes 
and libraries producing their own online materials or making their own 
search forms. Many of them are unaware of 
current accessibility standards.

Standards can seem difficult to apply. But really, they are easy to
follow if we make the guidelines clear enough so that everyone can understand 
and use them.

Remember that new technologies are taken into use all the time. We will
be constantly facing new challenges to make them accessible, but they will
also bring new possibilities with them.

\section{Don't Forget Copyright Law and License Agreements}

There is one last thing that you need to be aware of -- don't forget about
copyright. It is not a given fact that a library can freely distribute electronic material 
 to a patron who could then read it on a personal computer or some other 
device. 

The copyright laws in different countries vary surprisingly on this
point. Moreover, even when the right to access is written into a law, thus making special
exceptions to copyright for disabled persons, a license 
agreement between a library and a publisher might take this right away for particular electronic materials or products. A publisher or a consortium will not allow you to do things that
are not specifically stated in the signed agreement. Please always remember to 
check the accessibility options in agreements you sign.

To give an example, the current Finnish National Electronic Library (FinElib)
consortium agreement with Elsevier specifies that  ``Coursepacks in 
nonelectronic, non-print perceptible form (e.g. Braille) may be offered for [the] visually 
impaired.''\footnote{\texttt{http://wiki.helsinki.fi/display/FinELib/Freedom\#Freedom-inEnglish}} 
This is not, however, how visually impaired users
would like to use the materials. This is a standard clause that should be
modified to meet real needs.  Unfortunately, when the consortium was formed, this
clause did not receive the proper attention it should have.

\section{An Afterthought}

Practically everyone who lives long enough has to face physical challenges
at some point. An astronomer who is able-bodied today could have accessibility
issues tomorrow. We cannot expect that she or he is willing to give up practicing science. In her essay \textit{The Blind Astronomer} \citep{Isakssont2002},
the New Zealand astronomer Tracy Farr eloquently describes the changes 
brought by the gradual loss of her vision. With a different approach to
looking at the research data, she can continue to access the universe: 

\begin{quote}
I am freeing myself from the fixedness of the seen.
With my mind open to the universe, I hear the heavens' ebb and flow as
music. It is the incomprehensibly wonderful revelation of music first heard
after only ever having seen black spots and lines on a white page. As my
ears open and my eyes close, I hear the planets dance.

\end{quote}

\vskip 0.1in
					    
\acknowledgements

The author has received the University of Helsinki Chancellor's travel 
grant for LISA VI.

\end{document}